\documentclass[pdflatex,sn-mathphys-num]{sn-jnl}% Math and Physical Sciences Numbered Reference Style
\usepackage{graphicx}%
\usepackage{multirow}%
\usepackage{amsmath,amssymb,amsfonts}%
\usepackage{amsthm}%
\usepackage{mathrsfs}%
\usepackage[title]{appendix}%
\usepackage{xcolor}%
\usepackage{textcomp}%
\usepackage{manyfoot}%
\usepackage{booktabs}%
\usepackage{algorithm}%
\usepackage{algorithmicx}%
\usepackage{algpseudocode}%
\usepackage{listings}%
\theoremstyle{thmstyleone}%
\theoremstyle{thmstyletwo}%

\newcommand{\methodname}{BoomHQ}
\newcommand{\eg}{\textit{e.g.,}~}
\newcommand{\stitle}[1]{\vspace{1ex}\noindent{\bf #1}}

\theoremstyle{thmstylethree}%

\begin{document}

\title[Article Title]{BoomHQ: Learning to Boost Multiple Hybrid Queries on Vector DBMSs}


% \author*[1,2]{\fnm{First} \sur{Author}}\email{iauthor@gmail.com}



% \affil*[1]{\orgdiv{Department}, \orgname{Organization}, \orgaddress{\street{Street}, \city{City}, \postcode{100190}, \state{State}, \country{Country}}}

% \affil[2]{\orgdiv{Department}, \orgname{Organization}, \orgaddress{\street{Street}, \city{City}, \postcode{10587}, \state{State}, \country{Country}}}

% \affil[3]{\orgdiv{Department}, \orgname{Organization}, \orgaddress{\street{Street}, \city{City}, \postcode{610101}, \state{State}, \country{Country}}}

\author[1]{\fnm{Ermu} \sur{Qiu}}\email{qem@stu.pku.edu.cn}

\author[1]{\fnm{Tianyi} \sur{Chen}}\email{tianyichen@stu.pku.edu.cn}

\author*[1]{\fnm{Jun} \sur{Gao}}\email{gaojun@pku.edu.cn}

\author[2]{\fnm{Xing} \sur{Wei}}\email{wei.xing6@zte.com.cn}

\author[2]{\fnm{Yaofeng} \sur{Tu}}\email{tu.yaofeng@zte.com.cn}

\author[2]{\fnm{Yinjun} \sur{Han}}\email{han.yinjun@zte.com.cn}

\author[2]{\fnm{Yang} \sur{Lin}}\email{lin.yang@zte.com.cn}

\affil[1]{\orgdiv{Key Laboratory of High Confidence Software Technologies, CS}, \orgname{Peking University}, \orgaddress{\street{No. 5 Yiheyuan Road, Haidian District}, \city{Beijing}, \postcode{100871}, \state{Beijing}, \country{China}}}

\affil[2]{\orgdiv{ZTE Corporation}, \orgaddress{\street{No. 50, Software Avenue, Yuhuatai District}, \city{Nanjing}, \postcode{210012}, \state{Jiangsu}, \country{China}}}


\abstract{
Hybrid queries, which combine vector nearest neighbor searches with scalar predicates, represent a fundamental challenge in managing vector databases. 
Existing methods often restrict the number of vector columns involved or the complexity of scalar predicates, thereby limiting their flexibility in handling diverse query patterns.
Moreover, these approaches typically do not fully leverage the correlations between scalar and vector attributes, or the distributional patterns observed from query vector neighborhoods.
To address these limitations, we introduce \methodname{}, a learning-based framework to \underline{boo}st \underline{m}ultiple \underline{h}ybrid \underline{q}ueries on vector DBMSs.
First, \methodname{} models the correlation between vector and scalar attributes using an autoencoder-based architecture, which is also friendly to data updates. 
Second, \methodname{} captures prevailing query patterns, particularly using estimated selectivity of scalar predicates within the neighborhood of a query vector. 
Guided by these two key features, \methodname{} predicts the execution hints and rewrites the original query into an optimized version. 
Furthermore, we extend well-known benchmarks by introducing vector and scalar data with inherent correlations to better evaluate query execution. 
Experimental results demonstrate that for multiple hybrid queries at specified recall thresholds, our method achieves a $2\times$ average and over $25\times$ peak speedup compared to the state-of-the-art.
Additionally, \methodname{} shows strong robustness against data updates and consistent optimization effectiveness across three representative vector database systems.
}

%%================================%%
%% Sample for structured abstract %%
%%================================%%




\maketitle

\section{Introduction}\label{sec1}

Vector retrieval is a fundamental requirement in a wide range of data-driven applications~\cite{indyk1998approximate, arya1998optimal, wang2021comprehensive, li2026efficient, zhu2026braveann}. 
To meet the growing demand for efficient similarity search, numerous vector indexing techniques have been developed~\cite{li2019approximate}.
Among these, HNSW~\cite{malkov2018efficient} has emerged as one of the most prominent and widely adopted approaches. 
Benchmarking efforts like ANN-Benchmarks~\cite{aumuller2020ann} have further accelerated the advancement of approximate nearest neighbor (ANN) algorithms by providing standardized evaluation frameworks. 
Correspondingly, many systems and products have emerged to support vector retrieval across different application scenarios. These include native vector databases such as Milvus~\cite{wang2021milvus}, Weaviate~\cite{weaviate} and Pinecone~\cite{pinecone}, which are designed specifically for vector data management and indexing, as well as vector-extended systems built upon existing infrastructures, such as Pgvector~\cite{pgvector}, OpenSearch~\cite{opensearch}, and Elasticsearch~\cite{elasticsearch}. Together, these technologies have made vector retrieval a core capability in modern data systems.

Existing vector retrieval systems perform well when handling single-vector queries. 
However, in practice, it is common to encounter scenarios where vectors are combined with scalar attributes~\cite{liang2024unify, gollapudi2023filtered} or where multiple vector distances need to be considered jointly~\cite{qiu2025liftus, ann_2_yin2025deg}.

\begin{figure}[h]
  \includegraphics[width=\columnwidth]{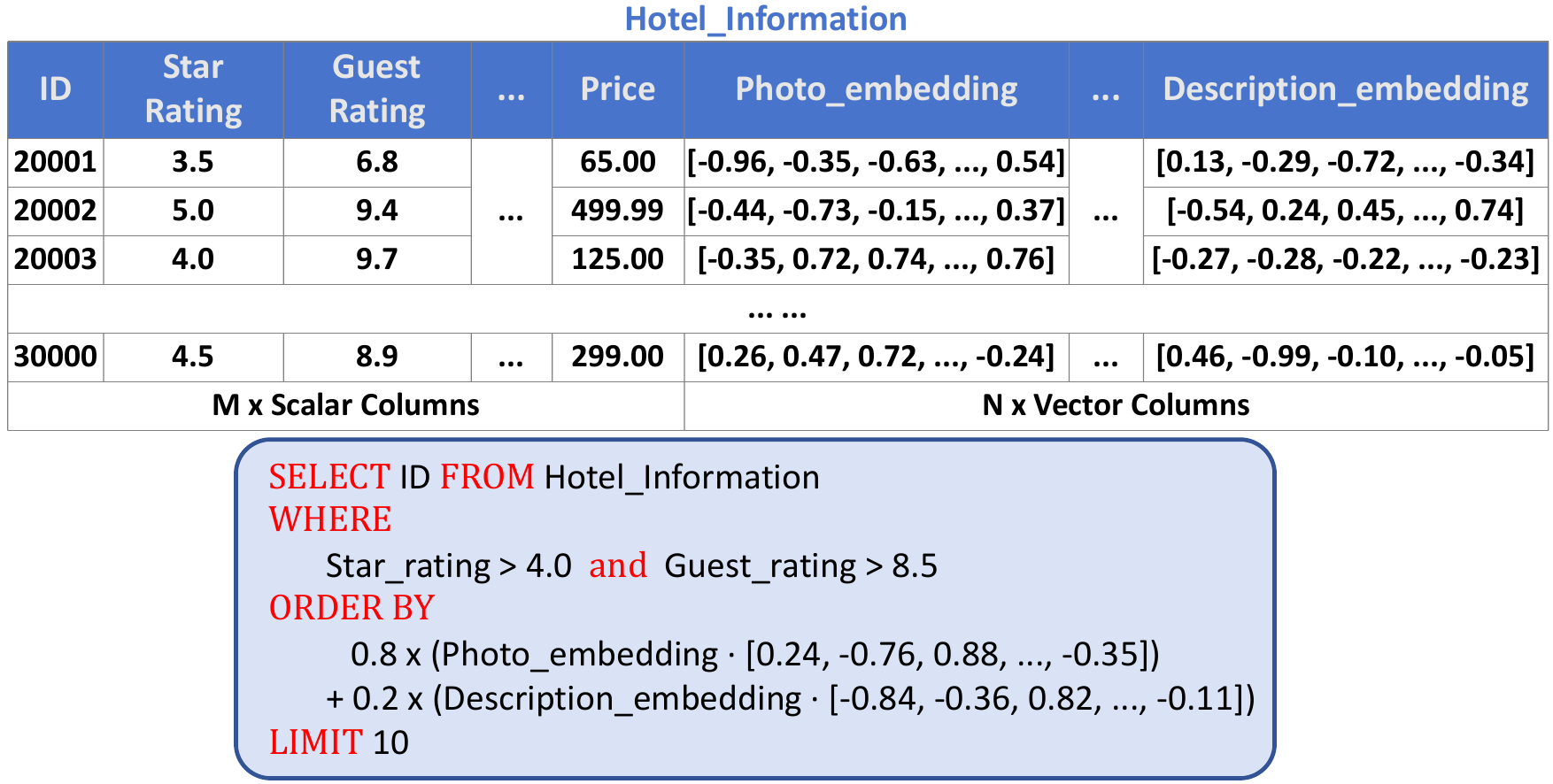}
  \caption{An example of a vector database and a weighted multi-vector hybrid query. There are $M$ scalar columns and $N$ vector columns in vector database. The query involves weighted vector nearest neighbors with scalar constraints.}
  \label{fig:example}
\end{figure}

We define queries that combine scalar filtering conditions with distance metrics over multiple vector columns as \textbf{Multiple Hybrid Queries (MHQ)}. 
Figure \ref{fig:example} illustrates an example of a MHQ in a hotel recommendation scenario, where users search for hotels with high ratings and similar visual and textual features. 
For instance, the following SQL query retrieves the top 10 hotels with a star rating above 4.0 and a guest rating above 8.5, ranking the results by a weighted combination of image and description similarities.
This query prioritizes hotels whose photos are visually similar to a given image while also considering the semantic relevance of their descriptions.

% Fig. \ref{fig:example} illustrates an example of a WMHQ. 
The example highlights the essential components of a MHQ, combining scalar predicates with weighted vector similarities. 
In the following, we provide a formal definition of MHQ.
The MHQ is operated on a table $\mathcal{T}$ containing $N$ vector columns and $M$ scalar columns.
For any tuple $o$ in table $\mathcal{T}$, it stores $N$ vectors $\{v_1, ..., v_N\}$ and $M$ scalars $\{s_1, ..., s_M\}$.
% We formally define such a query over a table $\mathcal{T}$ containing $M$ scalar columns $(s_1, ..., s_M)$ and $N$ vector columns $(v_1, ..., v_N)$. 
The query $\mathcal{Q}=<Q_S, Q_V, W_V>$ is composed of a set of scalar predicates $Q_S = \{p_1, ..., p_M\}$, a set of query vectors $Q_V = \{q_1, ..., q_N\}$, and a corresponding set of weights $W_V = \{w_1, ..., w_N\}$. 
% It first filters data using predicates $Q_S$ and then ranks the results based on a weighted composite score, defined as $Score(Q, o) = sum_{j=1 to N} w_j * sim(q_j, o.v_j)$, returning the top-k results with the highest scores.
After filtering the table by $Q_S$, all remaining objects $o$ are ranked by a composite score $Score(Q, o) = \sum_{i=1}^{N} w_i \operatorname{sim}(q_i, o.v_i)$, where $\operatorname{sim}(\cdot, \cdot)$ is a similarity function (\eg dot product or $L_2$ distance). The top-$k$ objects are returned as the final result.

Applying existing hybrid query methods to MHQ reveals a critical trade-off between effectiveness and efficiency.
For example, incorporating scalar predicates often lowers recall~\cite{liang2024unify, gollapudi2023filtered}, and multi-vector queries cannot directly leverage existing indexes~\cite{qiu2025liftus, ann_2_yin2025deg}.
To address these issues, we introduce a learning-based logical optimizer that predicts optimal execution strategies and parameters, enabling query rewriting to leverage existing index effectively on various vector DBMSs.

% In such cases, current systems often experience a significant degradation in performance.
Developing solutions for the MHQ problem is non-trivial due to the following key challenges.
% The challenges encountered by existing methods in multi-vector column hybrid queries are enumerated as follows.
First, a primary challenge is the highly dynamic correlation between vectors and scalars~\cite{fu2025vista}. 
Even in static environments, capturing the correlation between vector and scalar attributes is inherently challenging due to their heterogeneous representations and the lack of explicit semantic alignment~\cite{chronis2025filtered}.
This inherent static complexity is then significantly exacerbated by the highly dynamic nature of this correlation.
The continuous evolution means that any learned model of the vector-scalar relationship can quickly become outdated or invalid.
An optimization model trained on historical data may generate highly suboptimal query plans when faced with a new data distribution.
The overhead of model retraining is also a factor that cannot be ignored.
% This relationship is not static and constantly changes as new data is inserted into the database. 
% This continuous evolution means that some model of the vector-scalar relationship can quickly become outdated or invalid. 
% For instance, in an e-commerce context, a newly added trending product might suddenly establish a strong, unforeseen correlation with a specific demographic of users, perhaps those in a particular age group or geographic location. 
% This abrupt shift alters the existing correlation structure, rendering pre-computed indexes or cached query plans based on historical data less effective. 
% Consequently, the database must be able to adapt to these evolving patterns to maintain accurate and efficient query performance.
Therefore, maintaining an accurate and efficient modelling of vector–scalar correlations under dynamic workloads remains an open and fundamental challenge.

Second, the selectivity of scalar conditions in hybrid queries affects the efficiency and effectiveness of hybrid query algorithms~\cite{liang2024unify, wei2020analyticdb}.
HNSW~\cite{malkov2018efficient} and IVFflat~\cite{IVFflat} prioritize searching within the query vector’s immediate neighborhood. 
However, in hybrid queries, the tuples that satisfy the scalar constraints are not always located within this specific neighborhood. 
% For instance, a query might seek a product vector that is similar to a user's preference, but also requires the product to be available in a specific color or size. 
% The products with the correct color or size might be far from the nearest neighbors in the vector space. 
% This mismatch between the vector-based search area and the required data location forces the system to perform a wider, less targeted search, or even a full scan of the dataset. This significantly lowers retrieval efficiency.
For instance, an HNSW search can be inefficient if almost no points in a query vector's neighborhood satisfy the scalar conditions. 
The algorithm then fails to find enough qualified tuples. 
This failure prevents timely convergence and leads to high query latency. 
A filter-first approach that prioritizes the scalar conditions could yield better performance in such scenarios. 
In contrast, for queries with very high selectivity, prioritizing vector nearest search is more effective.
Consequently, the estimated value of selectivity is crucial for the selection of execution methods.
Moreover, due to the stochastic nature of the data, global selectivity may not accurately reflect the local scalar satisfaction rate in the neighborhood of query vector. 
However, the latter plays a more critical role in determining whether the query converges but remains underexplored.

Third, in the MHQ problem, the dynamically changing weights of multiple vectors pose a significant challenge to the nearest neighbor search.
Traditional vector retrieval research focuses on single-column query optimization. 
This approach is not well-suited for multi-vector queries with dynamic weights. 
In more complex scenario, every change in a weight can alter the overall score and the ranking of candidate tuples~\cite{ann_2_yin2025deg}. 
This dynamic nature makes the system less predictable and efficient. 
For example, in a personalized search application, user preferences are constantly changing. 
Each change to the weight of a vector column, such as a user's interest in ``photo'' versus ``description'' in Fig \ref{fig:example}, requires a different query execution strategy. 
A weight adjustment may not only reorder candidate vectors but also affect the selection of the most suitable index.
Consequently, a robust system must be capable of adapting its execution strategy dynamically.
% This is computationally expensive and creates significant performance bottlenecks. 
% A more flexible and adaptive approach is critically needed.

% Finally, there is a lack of suitable benchmarks. 
% Many of the existing datasets are not publicly available, which limits broader research and validation. 
Finally, there is a lack of publicly available datasets comprising both multiple vector columns and multiple scalar columns suitable for evaluating MHQ algorithms.
Due to privacy and security reasons, certain appropriate datasets have not been made publicly available~\cite{zhang2023vbase}.
% Furthermore, access to several relevant datasets remains restricted due to privacy and security concerns.
% Researchers cannot easily reproduce or compare results. 
Some public datasets have a simple schema or only contain a single vector column~\cite{HybridQueriesBenchmark, aumuller2020ann}.
This makes them insufficient for testing new algorithms for MHQ. 
Therefore, a new benchmark is needed. 
This benchmark should be publicly accessible and include multiple vector columns and scalar columns. 
Such a dataset would allow for a fair and standardized comparison of new methods.

To address the challenges previously mentioned, our basic idea is as follows:
(1) We adopt an autoencoder-based architecture to support the modeling of vector-scalar correlations in scenarios with data updates.
(2) To more accurately perceive the query vector's neighborhood, particularly in scenarios where scalar conditions exhibit skewed distributions, we introduce a local exploration of the query vector neighborhood in addition to the global selectivity estimation.
(3) We leverage the aforementioned information to implement learning-based logical optimization that including query strategy rewriting and parameter recommendation, thereby supporting various vector DBMSs.
(4) We construct a benchmark by designing methods to add new vector columns and scalar columns, which enables existing mainstream datasets to support the evaluation of MHQ.

The contributions of this paper are as follows:

\begin{itemize}

\item We propose \textbf{\methodname{}}, a learned optimizer to \underline{boo}st \underline{m}ultiple \underline{h}ybrid \underline{q}ueries that integrates the following 3 innovative components:

\begin{itemize}
% \item \textbf{Correlation-Aware Data Encoder for Evolving Datasets}: 
\item \textbf{Correlation-Aware Vector-Scalar Data Encoder}:
We propose a novel data encoder for the correlation-aware modeling of mixed vector-scalar data. It captures the intricate correlations between the vector and scalar. And its autoencoder-based architecture facilitates efficient data updates with minimal maintenance cost.

\item \textbf{Neighborhood Selectivity Enhanced Query Representation}: We develop a query encoder to guide the selection of query execution strategies. It efficiently models the query from local aspect (via neighborhood pre-probing) and global aspect (via selectivity estimation) to provide valuable insights with minimal time overhead.

% \item \textbf{A Model-Guided Strategy for MHQ}: 
\item \textbf{MHQ Rewriter Using Predicted Execution Strategies and Parameters}: 
% We develop a model that leverages the preceding information to rewrite execution strategies for MHQ and recommend appropriate parameter settings.
We develop a learning-based optimizer that performs logical query strategy rewriting and parameter recommendation, enabling support for various vector DBMSs.

\end{itemize}

\item We introduce a new, publicly available benchmark specifically designed for evaluating MHQ in vector database systems. 
This benchmark includes a rich schema and diverse data types, which provides a uniform tool for future research.

\item Experiment results show that \methodname{} achieves a 20\% QPS improvement on single-vector-column hybrid queries. For weighted multi-vector-column hybrid queries under a specified recall threshold, it delivers $2\times$ average and over $25\times$ peak speedup compared to the state-of-the-art. Furthermore, our method exhibits strong robustness against data updates and demonstrates consistent optimization effectiveness across diverse vector database systems.

\end{itemize}

\section{Related Work}
In this section, we first review the ANN and its graph-based methods, then introduce recent progress in hybrid query, and finally discuss vector database query optimization.

\subsection{ANN and Graph-based Methods}
ANN search aims to efficiently find the most similar vectors to a query in high-dimensional space~\cite{indyk1998approximate, arya1998optimal}. As an exact search is computationally prohibitive on large datasets, vector indexing techniques are employed to accelerate the process by trading a small amount of accuracy for a significant gain in speed. Major approaches include hashing-based\cite{indyk1998approximate}, clustering-based\cite{IVFflat}, quantization-based\cite{jegou2010product}, and graph-based~\cite{malkov2018efficient} methods.
Among them, graph-based methods demonstrate promising results and become a mainstream research direction recently~\cite{wang2021comprehensive}.

% \stitle{Hashing-based Methods.} Locality-Sensitive Hashing (LSH) uses hash functions designed to map similar vectors to the same bucket with high probability \cite{indyk1998approximate}. A query is performed by hashing the query vector and searching only within the corresponding bucket(s). Although LSH provides theoretical guarantees, achieving high recall often demands numerous hash tables, resulting in substantial memory usage.

% \stitle{Clustering-based Methods.} Clustering-based methods, such as the Inverted File (IVF) index, partition the vector space into clusters \cite{IVFflat}. Each vector is assigned to its nearest cluster centroid. At query time, the search is confined to the vectors within a subset of clusters closest to the query, controlled by the \textit{$n\_probe$} parameter. A larger \textit{$n\_probe$} increases accuracy at the cost of higher latency.

% \stitle{Quantization-based Methods.}
% Quantization-based methods like Product Quantization (PQ) compress vectors to reduce memory usage and accelerate distance calculations \cite{jegou2010product}. PQ works by dividing vectors into sub-vectors and quantizing each part separately, representing the original vector with a short code. This enables efficient approximate distance computations and is often combined with other structures like IVF to create a composite index.

% \stitle{Graph-based Methods.} 
% In recent years, graph-based methods have emerged as the state-of-the-art for ANN search, demonstrating superior performance across various benchmarks \cite{aumuller2020ann, wang2021comprehensive}. 
Graph-based methods construct a proximity graph~\cite{wang2021comprehensive} where nodes represent vectors and edges connect close neighbors. 
% The search becomes a graph traversal, starting from entry points and greedily navigating towards the query. 
The search process operates as a graph traversal that initiates at specific entry points and greedily advances toward the query vector.
Among these, Hierarchical Navigable Small World (HNSW) is particularly popular for its excellent performance \cite{malkov2018efficient}. HNSW introduces a multi-layer, hierarchical structure to Navigable Small World (NSW) graphs~\cite{nsw_malkov2012scalable, nsw_malkov2014approximate, nsw_ponomarenko2011approximate}. Top layers contain sparse graphs with long-range links for coarse navigation, while bottom layers have denser graphs with short-range links for fine-grained search. The search begins at a top-layer entry point, greedily finds the closest node, and uses it as the entry point for the layer below, repeating until the bottom layer is reached for a final, thorough search.
This efficient implementation mechanism underpins the widespread use of HNSW in modern vector DBMSs~\cite{pgvector, wang2021milvus, opensearch}.

Optimizations for HNSW can be categorized into two types: underlying graph structure optimization and external parameter control optimization.
Similar to HNSW, many methods achieve stronger ANN indexing by optimizing the underlying graph structure, such as DiskANN~\cite{jayaram2019diskann}, HM-ANN~\cite{ren2020hm}, and NSG~\cite{fu2019fast}. 
While these studies primarily focus on structural optimization, we observe that the performance of HNSW is also highly sensitive to parameter settings.
For instance, $ef\_search$ controls the size of the dynamic candidate list during search, directly trading off between recall and latency.
Furthermore, there are also parameters such as $iterative\_scan$ and $max\_scan\_tuples$ that can be adjusted~\cite{pgvector}.
There remains considerable room for exploring parameter optimization in graph-based methods.

\subsection{Hybrid Query}
Hybrid query, which combines vector-based similarity search with traditional attribute-based filtering, has become a critical requirement. Two main approaches have emerged: extending vector-native indexes to handle scalar attributes, and integrating traditional scalar-first databases to incorporate vector search capabilities.

\stitle{Extending Vector Indexes with Attribute Filtering.}
An approach to accelerating hybrid queries is to extend existing proximity graph~\cite{wang2021comprehensive} indexes, such as HNSW~\cite{malkov2018efficient} or DiskANN~\cite{jayaram2019diskann}, to natively integrate attribute constraints.
This avoids the performance limitations of simple pre-filtering or post-filtering.
Several strategies accomplish this by partitioning or segmenting the graph based on attribute values, particularly for range filters.
This includes methods like SeRF~\cite{ann_5_zuo2024serf}, which compresses multiple range-specific indexes into one, UNIFY~\cite{liang2024unify}, which creates a unified hierarchical segment graph to dynamically select a filter strategy, and iRangeGraph~\cite{ann_4_xu2024irangegraph}, which dynamically improvises a query-specific graph from pre-computed elemental graphs.
For label-based or arbitrary predicates, techniques either modify the graph construction process to be ``filter-aware'', such as in Filtered-DiskANN~\cite{gollapudi2023filtered} and UNG~\cite{ann_3_cai2024navigating}, or, like ACORN~\cite{ann_6_patel2024acorn}, use a predicate-agnostic index but perform an intelligent sub-graph traversal at query time.

\stitle{Integrating Scalar Databases with Vector Search.}
The alternative approach is to augment traditional scalar databases with vector search capabilities. This allows users to leverage existing database systems for hybrid queries. 
% A popular example is pgvector, an extension for PostgreSQL that adds a new vector data type and indexing methods (such as IVF and HNSW) \cite{pgvector}, enabling developers to combine SQL-based filtering with ANN search in a single query. 
Pgvector\cite{pgvector} is an extension for PostgreSQL that adds a new vector data type and indexing methods (such as IVF and HNSW), enabling developers to combine SQL-based filtering with ANN search in a single query. 
VBase\cite{zhang2023vbase} also expands PostgreSQL by leveraging the property of ``relaxed monotonicity'' to decide whether continuing the search from the current node is likely to yield better candidates.
% Another prominent example is Elasticsearch, which has integrated dense vector fields and approximate k-NN search into its distributed search engine \cite{elasticsearch}. 
Milvus~\cite{wang2021milvus}, OpenSearch~\cite{opensearch}, and ElasticSearch~\cite{elasticsearch} provide integrated solutions for combining scalar database management with high-performance vector search capabilities.
This allows for powerful hybrid queries that combine traditional full-text search, structured filters, and vector similarity search, fusing relevance scores from different query clauses to provide more comprehensive results.

\subsection{Vector Database Query Optimization}
% Another line of work treats the vector search process as a black box and applies external models or policies to optimize its performance without modifying the core index structure. These methods primarily focus on selecting execution strategies or dynamically adjusting search parameters and termination conditions on a per-query basis.
Beyond internal algorithmic tuning, we review optimization strategies applied externally to vector search, focusing on execution strategy selection and learned parameter tuning.

\stitle{Execution Strategy Selection.}
One approach is to introduce a higher-level query optimizer that selects the most efficient execution strategy for a given query. For hybrid queries, this often involves deciding the order of vector search and attribute filtering. For example, Alibaba's AnalyticDB for Vector (ADBV) employs a cost-based optimizer that chooses between a pre-filtering (attribute-first) and a post-filtering (vector-first) strategy \cite{wei2020analyticdb}. The decision is guided by a learned model that estimates the selectivity of the attribute filter, which allows the system to dynamically select the cheaper execution plan based on the query's specific characteristics.

\stitle{Learned Parameter Tunning.}
This approach aims to predict optimal search parameters for each query. One line of research focuses on cardinality estimation for similarity queries. For instance, \cite{sun2021learned} uses a deep neural network to estimate the number of objects within a given distance threshold by learning distributions at a fragment level and composing them. Other works predict the optimal $n\_probe$ value for IVF-style indexes. PCE-Net uses an encoder-decoder architecture to estimate the minimum probing cardinality needed for a given query \cite{zheng2023learned}. Similarly, LIRA learns a query-aware model that outputs a probability for probing each cluster, which allows for an  adaptive and fine-grained search strategy~\cite{zeng2025lira}.

\section{methodology}

In this section, we first provide an overall framework of our proposed \methodname{}, a novel system designed to address the challenges of efficient MHQ processing. 
% The core objective of \methodname{} is to automatically select the optimal execution plan by deeply understanding the inherent correlations between vector and scalar attributes. 
\methodname{} is composed of three key stages: (1) a correlation-aware vector-scalar data encoder based on an autoencoder to learn the joint representation of query-independent data, (2) a query encoder that leverages neighborhood pre-probing and global selectivity estimation, and (3) an MHQ rewriter using predicted execution strategies and parameters.

\begin{figure*}[t]
  \includegraphics[width=\textwidth, trim={75mm 30mm 0mm 63mm}, clip]{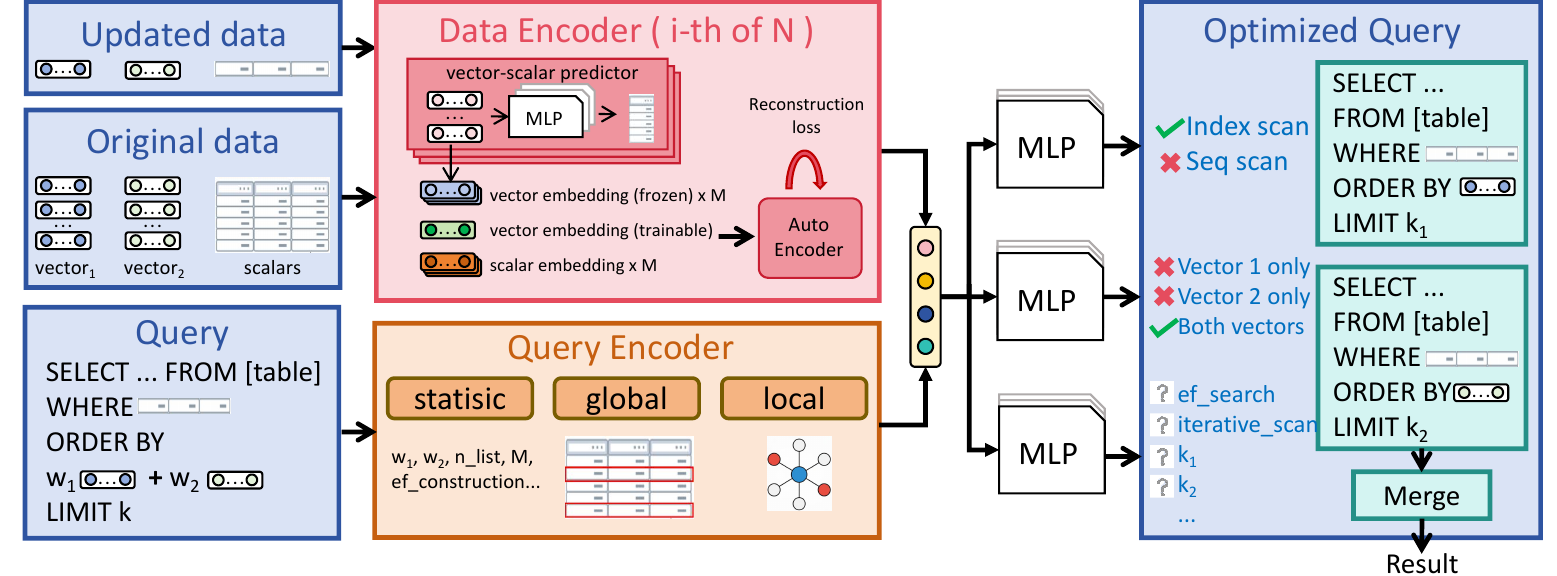}
  \caption{Architecture of \methodname.  }
  \label{fig:architecture}
\end{figure*}

\subsection{Framework}
In this paper, we choose to employ a logical optimization approach for the MHQ problem. 
This approach involves the recommendation and tuning of relevant execution strategies and their corresponding parameters.

The search space of the logical optimization to MHQ problem is as follows.
The selection of execution strategies encompasses sequential scan and index scan.
For all HNSW-based hybrid queries, the optimization process includes the recommendation of the parameters $ef\_search$, $max\_scan\_tuples$, and $iterative\_scan$ .
In the context of MHQ, which involves multiple vector columns, it is also necessary to recommend the candidate set size for each respective column.
Furthermore, to accommodate scenarios with highly skewed weights, we also introduce an alternative strategy that performs the ANN search on a single vector index only.

To accurately recommend these aforementioned values, we should gather relevant features from both the data and query.
As illustrated in Figure~\ref{fig:architecture}, \methodname{} consists of two major components, data encoder and query encoder.
The data encoder (Section 3.2) takes as input only the original data, which includes multiple vector columns and scalar attributes.
To handle dynamic updates, new data are first stored in a buffer and incrementally integrated into the encoding process, which ensures the model remains effective under evolving data distributions.
The query encoder (Section 3.3) analyzes incoming SQL-style queries. It parses each query to extract a tripartite set of features: statistics, global selectivity estimation result, and local neighborhood pre-probing information.
Information from both encoders is finally fed into MHQ rewriter (Section 3.4) to predict execution strategy and parameters.

\subsection{Correlation-Aware Vector-Scalar Data Encoder}
In many real-world applications, vector and scalar attributes are not independent. For instance, in an e-commerce dataset, the vector embedding of a product image is often correlated with scalar attributes like price, brand, or category. An image of a luxury watch (vector) is inherently linked to a high price and a specific brand name (scalar). Vector databases should account for these correlations to perform efficient hybrid queries. When a query's vector-scalar relationship aligns with the learned data distribution (\eg searching text embeddings for ``deep learning'' papers with a recent publication year), an HNSW-based search can rapidly converge on qualified results within the vector's local neighborhood. Conversely, a query with an anomalous pairing (\eg ``deep learning'' papers before 1980) may force the search algorithm to traverse distant, dissimilar nodes in the graph, significantly expanding the search space and degrading performance.

Furthermore, with the insertion of new data, the underlying vector-scalar correlations may shift. An effective model should not only capture the initial relationships but also adapt to new patterns efficiently without requiring a full-scale, costly retraining process on the entire dataset. To address these challenges, we propose a correlation-aware model based on a specialized autoencoder architecture.

Our model is designed to learn the joint distribution of vector-scalar pairs, with support for incremental data insertion.
To achieve this, we encode each vector by taking into account its correlation with the corresponding scalar. 
And we employ an autoencoder architecture to jointly model these vector-scalar pairs, a structure that facilitates incremental updates. 
We describe each step in detail below.

\stitle{Vector Encoding.} 
We consider a scenario where a set of $N$ vectors $v_1, v_2, ..., v_N$ are stored in the database. 
Without loss of generality, the following discussion will focus on the $i$-th vector $v_i$.
Instead of using the raw vector directly, we project $v_i$ using a dual-pathway network. 
This projection is composed of two components: one that considers the relationship between the scalar $s_1, s_2, ..., s_M$ and $v_i$ (frozen), the other that extracts the essential information from $v_i$ itself (trainable).

For each vector $v_i$, in the frozen part, we establish a set of $M$ corresponding neural networks, denoted as $f_{i,j}$ for $j=1, \dots, M$. 
Each network, architected as a MLP, is independently trained to predict a specific scalar target $s_j$ from the input vector $v_i$. 
The training objective is to find the optimal model parameters $\theta_{i,j}^*$ that minimize a loss function $\mathcal{L}_{i,j}$, which can be tailored for either a regression or a classification task. 
This process is formally expressed as:$$\theta_{i,j}^* = \arg\min_{\theta_{i,j}} \mathcal{L}_{i,j}(f_{i,j}(v_i; \theta_{i,j}), s_j)$$
Upon completion of this training phase, the parameters of each network $f_{i,j}$ are frozen, denoted by $f_{frozen_{i,j}}$. 
Consequently, they are not updated during subsequent backpropagation stages of the main model. 
These networks are exclusively retrained when new data is introduced into the system.
% The frozen network is tasked with explicitly capturing the predictive relationship between the vector v and its associated scalar attributes s. 
% Its parameters are learned in a preliminary supervised task to predict s from v. 
This process embeds scalar-relevant information directly into the vector's representation. 

In the trainable part, we employ $f_{trainable_i}$ to transform the original $v_i$ into a dense, task-relevant embedding.
This process effectively extracts the essential information and learns a high-level feature representation required for the following  recommendation of the optimal execution method.
Unlike the frozen networks, the parameters of this MLP are fully trainable. 
They are continuously updated via backpropagation during the end-to-end training of the overall model.

% The trainable network learns to capture additional latent features from the vector.

% Let $v \in R^d$ be the input vector and $s_1, s_2, ..., s_M$ be the $M$ scalar attributes. 
The final projected vector, $E_{v_i}$, is formulated as the concatenation of the $M$ frozen networks and the trainable network:$$E_{v_i} = [\big\|_{j=1}^{M} f_{frozen_{i,j}}(v_i); f_{trainable}(v_i)]$$

Here, $\big\|$ and $[;]$ denote the concatenation operator. 
This dual-pathway approach ensures that the resulting embedding retains both a latent representation of the vector and explicit features correlated with each specific scalar attribute.

\stitle{Scalar Encoding.} Scalar attributes are encoded into a numerical format suitable for model ingestion. For categorical scalars with a manageable number of unique values, we employ standard one-hot encoding. 
% For continuous numerical values or high-cardinality categorical features, we first discretize the values into a predefined number of bins, allowing the value range of each bin to be transformed into a one-hot vector representation.
For continuous numerical values or high-cardinality categorical features, we first discretize the values into a predefined number of bins and then transform the value range of each bin into a one-hot vector representation.
Using these 2 methods, each scalar $s_j$ in the table $\mathcal{T}$ is then represented by a one-hot vector $g_j(s_j)$. 
This method concatenates the one-hot vectors corresponding to all scalars to obtain a unified representation $E_s=\big\|_{j=1}^{M} g_j(s_j)$.

\stitle{Autoencoder Training.}
The $i$-th encoded vectors and the representation of all scalars are concatenated to form an embedding: $E_i= [E_{v_i}; E_s] $. This combined embedding serves as the input to a standard autoencoder. The autoencoder is trained to minimize the reconstruction error between the input $E_i$ and the output $E'_i$, thereby learning the underlying manifold of common vector-scalar pairs in the dataset.

\stitle{Query Phase.}
During the query phase, the trained autoencoder is used to evaluate the conformity of a query's vector-scalar pair with the learned data distribution. Given a query vector $q_i$ corresponds to $i$-th column and scalar predicates $Q_s$, we apply the same transformation and encoding pipeline to generate the combined embedding $E_{q_i}$. This embedding is then passed through the autoencoder to produce a reconstructed version, $E'_{q_i}$.

The reconstruction error, calculated as the mean squared error between the original and reconstructed embeddings $\varepsilon_{recon_i} = \| E_{q_i} - E'_{q_i} \|^2$, serves as a powerful deviation score. A low reconstruction error signifies that the queried vector-scalar combination is well-represented in the training data and conforms to the learned correlations. 
% In contrast, a high error indicates an anomalous pairing, suggesting that vectors similar to $v_{q_i}$ are highly unlikely to satisfy the scalar condition. 
In contrast, a high error indicates an anomalous pairing, which suggests that vectors similar to $v_{q_i}$ are highly unlikely to satisfy the scalar condition. 
This anomaly score is a critical metric for the query execution engine, which enables it to estimate search difficulty and adjust its strategy accordingly. For instance, by broadening the initial search ranges for queries with high reconstruction error.

\stitle{Incremental Model Updates.}
A significant advantage of our approach is its ability to adapt to new data efficiently. When a new batch of data is inserted into the database, we do not need to retrain the autoencoder on the entire dataset. Instead, we can perform incremental fine-tuning by continuing the training process using only the newly added data. This method allows the model to quickly learn evolving vector-scalar correlations and ensures its representation remains current without the prohibitive computational overhead of a full retraining cycle. This makes the model practical for real-world database environments that are constantly evolving.

% \subsection{Neighborhood Pre-probing and Global Selectivity Estimation}
\subsection{Neighborhood Selectivity Enhanced Query Representation}

In the preceding section, we detailed a model for capturing the correlation between vector and scalar attributes. 
% While this correlation information is vital for understanding the data distribution, another critical aspect of optimizing hybrid query performance is the dynamic selection of the most efficient execution strategy. 
In addition to the data correlations, query-relevant features are also of critical importance.
% This section introduces a approach that combines local neighborhood analysis with global data statistics to guide the query execution plan.
As shown in Figure~\ref{fig:architecture}, we categorize the query-relevant features into three types:
(1) \textbf{Statistic}-related information, including the data explicitly presented in the query (\eg weights $w_i$, number of returned candidates $k$).
(2) Information jointly determined by vectors and scalars, which denotes the satisfaction rate of scalar conditions within the \textbf{local} neighborhood of query vectors.
(3) Information associated with scalar conditions, such as the \textbf{global} selectivity.

The choice of an execution strategy for a hybrid query (primarily between an index-first scan and a filter-first (sequential) scan) is pivotal. When the scalar predicates are satisfied by a high density of vectors in the immediate neighborhood of the query vector, an ANN index scan can converge rapidly to a satisfactory result. In this scenario, a sequential scan would be needlessly inefficient. However, the converse is also true. 
If the nearest neighbors returned by an index search consistently fail to meet the scalar predicate, algorithms like HNSW or IVF may be forced to explore an ever-expanding search space, which results in high latency~\cite{malkov2018efficient, IVFflat}. 
Intervening with a early termination condition (\eg a maximum number of scanned tuples) may avoid the latency but at the cost of recall, as an insufficient number of qualifying results might be found. In such cases, a sequential scan, aided by indexes on scalar columns, could provide superior accuracy within an acceptable latency. Therefore, the optimal query strategy is contingent upon both the local data distribution around the query vector and the global selectivity of the scalar filters. Optimizing this choice is key to enhancing the efficiency and accuracy of hybrid queries.

Building on this, the optimal tuning of search-time parameters for ANN algorithms is also intrinsically linked to these factors. Parameters such as $ef\_search$ in HNSW, which controls the size of the dynamic candidate list during search, directly impact the trade-off between search speed and accuracy. A static, one-size-fits-all parameter setting is suboptimal. For a query whose neighborhood is dense with qualifying data points, a smaller $ef\_search$ is sufficient. Conversely, for a query in a sparse region, a larger $ef\_search$ is necessary to find enough candidates that satisfy the scalar predicate. It thus becomes clear that an effective query optimizer requires a mechanism to estimate local and global data characteristics to dynamically recommend or adjust these crucial search parameters on a per-query basis.

\stitle{Neighborhood Pre-probing.} A primary challenge in index-first hybrid search is the risk of poor convergence when the local satisfaction rate for scalar predicates is low. Pure ANN searches, however, do not suffer from this issue and typically converge very quickly. We leverage this property to design a lightweight pre-probing mechanism to explore the query's neighborhood before committing to a full execution plan.

Upon retrieving the initial set of nearest neighbors, the system evaluates their scalar attributes against the query's specified predicates. From this evaluation, it computes a local satisfaction rate, which is formally defined as the number of probed neighbors that satisfy the scalar conditions divided by the total number of neighbors retrieved in the probe. This simple yet powerful metric serves as a direct signal. It quantifies the density of qualifying data points within the immediate neighborhood of the query vector and effectively summarizes a complex local data landscape into a single value.

A critical aspect of our methodology is how this local satisfaction rate is leveraged to guide query optimization. A conventional approach might employ a series of hard-coded, rule-based heuristics to make a decision, for example, triggering an index scan if the rate exceeds a predefined threshold. However, such static rules lack robustness and struggle to capture the complex dynamics of query execution. The optimal threshold can vary significantly based on other factors, such as the global selectivity of the filter and the underlying index structure.

Therefore, instead of relying on rigid heuristics, our framework treats the local satisfaction rate as a key input feature for a predictive model, which will be detailed in a subsequent section. This learned approach allows the query optimizer to make a far more nuanced decision. The model considers this crucial piece of local evidence not in isolation, but in conjunction with other signals, such as the global selectivity estimates and other query-specific characteristics. By modeling the complex relationships between these inputs, the optimizer  evaluates query context to derive the suitable execution plan and ensure an adaptive processing strategy.

\stitle{Global Selectivity Estimation.} While local pre-probing assesses the neighborhood immediately surrounding a query, global selectivity estimation provides a complementary, macro-level view of the filtering power of scalar predicates. The selectivity of a filter, which is the fraction of the tuple that satisfies the predicate, significantly influences the optimal choice for a query execution strategy. When a filter is highly selective and matches only a small portion of the tuples, a filter-first approach becomes highly advantageous. This strategy employs an inexpensive scalar filter to reduce the candidate set, and the subsequent vector search operates only on the filtered records. Conversely, when a filter matches a large fraction of the data, an index-first approach is more efficient. In this scenario, the exploration of the query vector’s neighborhood exhibits a high probability of encountering tuples that satisfy the scalar predicate constraints, thus accelerating query convergence.

To provide this crucial information without performing a costly full data scan at query time, we employ a histogram-based estimation algorithm. This technique relies on statistics pre-computed and stored for each scalar column, typically during data ingestion or as an offline background process. For numerical columns, the system partitions the entire range of values into a series of discrete bins and maintains a count of the records that fall into each. For categorical columns, the algorithm builds a frequency map that stores the total count for each distinct category. This preprocessing step creates a compact statistical summary of the data distribution for each attribute.

At query stage, these pre-computed histograms allow for a nearly instantaneous estimation of a filter's selectivity. 
On receiving a query with scalar predicates, the system parses the conditions and consults the corresponding histograms.
During the offline preprocessing stage, after the initial bin counts are established, a prefix sum array is generated for the histogram. Each entry in this array stores the cumulative count of all data points up to the upper boundary of the corresponding bin. 
For example, when a query with scalar predicate $price < 50$ arrives, the system performs a binary search over the bin boundaries to efficiently locate the bin that contains the predicate's threshold. The value in the pre-computed prefix sum array at that bin's index then provides a very close estimate of the total count. While this method can introduce a small margin of error when the query's threshold does not align perfectly with a bin boundary, it offers an extremely fast and effective mechanism for estimating the selectivity of range queries. When a query involves multiple conjunctive predicates, their joint selectivity is estimated by multiplying the individual selectivities together, a standard practice that assumes independence between the attributes. While this assumption may not always hold perfectly, it provides a fast and effective approximation that serves as a useful signal for
guiding the query optimizer's strategic decisions.

% \stitle{Synergistic Nature of Probing and Estimation.} 
% The neighborhood pre-probing and global selectivity estimation modules serve complementary roles and jointly provide a robust two-level view of the data.
% Global selectivity estimation alone is insufficient because its aggregate statistical view may not hold true for the specific local region of the vector space where a query falls. 
% The underlying data is often non-uniformly distributed, which produces localized areas with scalar attribute that deviate significantly from the global average.
% A query optimizer that relies only on a global estimate may overlook local deviations and select a suboptimal execution strategy.

% Conversely, relying solely on neighborhood pre-probing can also be deceptive. A small, fast probe might luckily find a few neighbors that satisfy a globally rare predicate, wrongly suggesting that the neighborhood is dense with qualifying results. Alternatively, a probe might fail to find any matches in the immediate vicinity, suggesting a sequential scan, when the global selectivity is high and a slightly more exhaustive index search (\eg with a higher $ef\_search$) would have been highly effective. Therefore, our framework uses both signals: global selectivity sets a strong baseline for the initial strategy recommendation, while neighborhood pre-probing acts as a dynamic verification and correction mechanism, confirming or overriding the initial plan based on the concrete, local evidence around the query vector. This synergy enables a far more adaptive and resilient query optimization strategy.

\subsection{MHQ Rewriter Using Predicted Execution Modes and Parameters}
The previous sections detailed a correlation-aware vector–scalar data encoder and a query representation enhanced by selectivity signals at both global and neighborhood levels. 
These analyses provide vital information about the data and query. 
This subsection explains how we use this hint in our two-phase query processing framework. The framework aims to rewriter MHQ by selecting an efficient execution strategy and recommending optimal query parameters.

\stitle{Query Execution Strategy Prediction.}
To achieve low latency and high recall for HNSW, tuning query parameters like $ef\_search$ and the $iterative\_scan$ is essential. We developed a prediction model to automate this process. The model determines the best parameters based on the specific characteristics of a given query.

The inputs to our model are designed to capture a comprehensive view of the data and the query. 
Let the input feature vector be denoted by $X_{in}$. It is constructed as follows:
% $X_{in} = [V_{qsample} \oplus S_{enc} \oplus I_{params} \oplus L_{recon} \oplus R_{probe} \oplus \sigma_{est}]$
$$X_{in} = [\big\|_{i=1}^{N} \varepsilon_{recon_i} ; S_{enc} ; E_{rec} ; R_{probe} ; \sigma_{est}]$$

Here, $\varepsilon_{recon_i}$ is the reconstruction loss of $i$-th vector column, $S_{enc}$ is an encoded representation of the scalar predicate information, and $E_{rec}$ represents the recall threshold expected by the user. Additionally, $R_{probe}$ contains the results from the neighborhood pre-probing stage, and $\sigma_{est}$ is the estimated global selectivity for the scalar predicate. 
% The symbol $\oplus$ denotes the concatenation operation.

For the model architecture, we employ a MLP for quick inference. This network is well-suited for learning complex, non-linear relationships between the input features and the optimal query parameters. The network consists of several fully connected hidden layers with ReLU activation functions, followed by an output layer that predicts the target parameters.

We train this model in a self-supervised manner. This approach is necessary because obtaining a large, labeled dataset of optimal query parameters is often impractical. For self-supervised training, we generate a large set of random queries. We execute each query with various parameter configurations and measure the resulting performance (\eg latency and recall). 
We define the model’s optimization target as the parameter configuration for each query that yields the minimum latency while satisfying the specified recall threshold.
% The model is then trained to predict the parameters that minimize a loss function, which is designed to balance execution time and search accuracy.

\stitle{Two-Phase Strategy for Multi-Vector Queries.}
For hybrid queries involving multiple vector columns, we introduce a two-phase processing strategy. To illustrate, a typical weighted hybrid query over two vector columns can be expressed in SQL as follows:

\begin{quote}
\ttfamily
SELECT id 
FROM   X 
WHERE $Q_s$
\\
ORDER BY $w_1 * (vec_1 \leftrightarrow q_1) + w_2 * (vec_2 \leftrightarrow q_2)$ \
LIMIT $k$;
\end{quote}

% Here, $\texttt{w_1}$ and $\texttt{w_2}$ are weights, $\texttt{vec_1}$ and $\texttt{vec_2}$ are the vector columns, and $\texttt{q_1}$ and $\texttt{q_2}$ are the corresponding query vectors.

Here, $Q_s$ is the scalar predicate, $w_1$ and $w_2$ are weights, $vec_1$ and $vec_2$ are the vector column names of $v_1$ and $v_2$, $q_1$ and $q_2$ are the corresponding query vectors, and $\leftrightarrow$ represents a set of vector distance metric functions.

\paragraph{Phase 1: Execution Plan Selection}
The first phase involves a model that selects the most efficient execution plan. The inputs to this model are the same as the parameter prediction model, with the addition of the query weights, $w_1$ and $w_2$. The model's primary output is a decision between two execution methods:

\begin{enumerate}
\item \textbf{Full Table Scan:} This method is chosen when scalar predicates have high filtering effectiveness or when vector index scans are predicted to be inefficient. 
\item \textbf{Vector Index Scan:} This method is favored for queries where scalar predicates fail to significantly reduce the candidate set, which means that leveraging the vector index is advantageous.
\end{enumerate}

If the ``Full Table Scan'' is chosen, the system proceeds directly with a full scan (or leverages available indexes on scalar columns, such as B-tree, bitmap, or hash indexes).
If the ``Vector Index Scan'' is chosen, the model performs the following phase 2. 

\paragraph{Phase 2: SQL Rewriting and Parameter Recommendation}
% It predicts an initial, larger limit ($k' > k$) for each vector column. It also predicts the optimal $\texttt{ef\_search}$ values and the $\texttt{iterative\_scan}$ method for each index scan.
% This phase is executed only if the "Vector Index Scan" plan was selected in Phase 1. 
% The individual index scans produce two initial sets of candidate results. These candidates are then merged. The final ranking is performed on this merged set by applying the original \texttt{ORDER BY} clause. This step accurately computes the combined distance and returns the final top-k results to the user.
For hybrid queries involving multiple vector columns, pre-calculating a combined index based on the weighted sum of the vectors is impractical. 
Such an approach would require rebuilding the index whenever the weights are updated, which is computationally expensive and introduces unacceptable latency for real-time applications.

To address this challenge, we propose a method that consists SQL rewriting and parameter recommendation. Instead of attempting a single complex search, we decompose the query into simpler subproblems. We then use a learning-based model to predict the optimal execution parameters for each subproblem, ensuring efficient execution.

Our method begins by rewriting the original multi-vector hybrid query into a set of individual subqueries. Each subquery performs a nearest neighbor search on a single vector column while still applying the original scalar conditions. This decomposition allows us to leverage existing single-vector indexes effectively.
A subquery for a single vector column $vec_i$ would be structured as follows:

\begin{quote}
\ttfamily
SELECT id, $vec_i \leftrightarrow q_i $ AS $d_i$ FROM X
\\
WHERE $Q_s$ ORDER BY $d_i$
LIMIT $k_i$;
\end{quote}

This subquery retrieves the $k_i$ nearest neighbors from the column $vec_i$ that also satisfy the specified scalar predicate. The process is repeated for each vector column involved in the original query.

The performance of each subquery heavily depends on its execution parameters. 
We use a model to recommend optimal values for key parameters like the result count $k_i$, $ef\_search$, and $iterative\_scan$.
The choice of $k_i$ is critical and is influenced by factors mentioned previously, such as vector-scalar correlation and neighborhood information. 
More importantly, its value is highly dependent on the dynamic weight $w_i$ of its vector column. 
Our model builds upon previous task by incorporating $w_i$ as an additional key input feature, which enables the system to adapt its strategy for each query.
In addition to $k_i$, the model also predicts $ef\_search$ and $iterative\_scan$ for each subquery, which allows us to fine-tune the execution plan for better performance.
Finally, the results returned by all subqueries are merged and subsequently re-ranked to generate the result of the original MHQ.

\subsection{Complexity Analysis}
This subsection analyzes the complexity of our proposed method. 
% Our approach operates as an external module, treating the core database as a black box. 
Our approach operates as an external module, agnostic to the database's core kernel.
Therefore, the analysis focuses on the additional time and space complexity introduced by our components.

The additional time complexity is divided into 2 primary sources. 
We first consider the costs on the data side, which involve the initial setup and subsequent data updates. For the initialization phase, our method operates on a sampled subset of the data, which significantly reduces the computational cost. More specifically, the feature engineering for each data item consists of two parts. One part of the process involves mapping $N$ vector $v_1, v_2, ..., v_N$ to a high-dimensional embedding via $M$ frozen and 1 trainable neural network transformation.
We denote $D$ as the time consumption of matrix multiplication.
The computational cost of this mapping for $N$ vector column of 1 sampled tuple, denoted as $c_v=N(M+1)D \ll \mathcal{N}$, is constant for a given model architecture. 
The other part transforms the scalar attributes $s_1, s_2, ..., s_M$ into an embedding. 
This is done by applying an encoding function to each attribute and concatenating the results. 
The cost for this scalar transformation, $c_s$, is also a small constant. 
The total time to process a single data item is therefore $c_v + c_s$, which we can define as a small constant $c$. These vector and scalar embeddings are then concatenated to train an autoencoder. Because these operations are performed on a representative sample, the overall time complexity for initialization is $O(c\mathcal{N})$, where $\mathcal{N}$ is the total number of items. This complexity is substantially lower than that of building indexes on the full dataset from scratch.

Our method also efficiently handles data updates. When new data arrives, we employ an incremental training strategy for the autoencoder. The model is fine-tuned using only the newly inserted data, instead of retraining on the entire dataset. This approach ensures that the system can adapt to data changes with minimal overhead. Consequently, the time complexity for processing updates is proportional to the number of new items, $\mathcal{M}$. This results in a time complexity of $O(c\mathcal{M})$ for update process.

Next, we analyze the time complexity on the query side. This refers to the overhead involved in determining an optimal execution plan and its associated parameters for a given query. This overhead is composed of two main parts. The first is the neighborhood probing module. This step performs a single, fast ANN search to retrieve a small set of candidate neighbors. It then checks the scalar predicate satisfaction rate within this set. The complexity of this module is dominated by the single ANN query, which is typically polylogarithmic with respect to $\mathcal{N}$. 
Consequently, neighborhood pre-probing exhibits the same time complexity as the subsequent query, and with a smaller constant factor since no scalar predicates are applied.
The second part is the global selectivity estimation. This module leverages pre-computed histograms for each scalar column. At query time, estimating the selectivity of a given predicate only requires a quick lookup in these histograms. This operation has a very low, near-constant time complexity.

Finally, we discuss the space complexity of our approach. The additional space overhead is primarily for storing the trained models and the pre-computed statistical information. This includes the vector mapping network, the autoencoder, and the histograms for each scalar attribute. 
% The size of these components depends on the model architecture and configuration, not the volume of the underlying data.
Therefore, our method exhibits constant space complexity, independent of the number of tuples stored in the database.

\section{Benchmark}
% In the era of large models, the demand for complex query capabilities has grown significantly. Modern applications increasingly require hybrid search, which combines semantic vector search with traditional structured scalar filtering. Despite this growing need, there is little standardized benchmark to evaluate the performance of different hybrid search solutions. A unified benchmark is crucial. It would allow researchers and engineers to fairly compare different systems, identify performance bottlenecks, and drive innovation in the field.

% Current benchmarks for vector and hybrid search face several challenges. 
Currently, the evaluation of MHQ performance faces many challenges.
First, there are very few publicly available datasets. 
Many existing benchmarks focus only on ANN search and lack corresponding scalar attributes. 
% Second, datasets that do include both data types, such as the Fungis dataset~\cite{HybridQueriesBenchmark}, often have a limited scale. This makes it difficult to test the performance of systems on large, industrial-sized workloads. 
% Second, the data schemas in these benchmarks are often too simple. 
Second, the schema structures in existing benchmarks remain fairly simple.
% They do not reflect the complexity of real-world applications. 
They typically contain only a limited number of scalar attributes that are applicable to MHQ.
Third, the query workloads are often not diverse enough to test various search conditions and filtering selectivities. These limitations make it hard to get a true measure of a system's capabilities.

To address these issues, we developed a new benchmark by creating large-scale hybrid datasets. We used two primary methods to construct our data.

\stitle{Augmenting Vector Datasets with Scalar Data.}
Standard vector datasets, like those in ann-benchmark, contain only vector information. Some recent work, such as~\cite{engels2024approximate}, added randomly generated scalar data to these vector datasets. However, in real-world scenarios, the vector and scalar attributes of an object are often correlated.

To create more realistic data, we generated correlated scalar attributes for existing vector datasets using three techniques:

\begin{itemize}

\item \textbf{Vector Clustering}: We first performed cluster analysis on the entire set of vectors. Each vector was then assigned to a specific cluster. The ID of the cluster (\eg Cluster 0, Cluster 1, Cluster 2) was used as a discrete scalar label for that vector. This method creates a categorical feature based on the vector's position in the embedding space.

\item \textbf{Random Hyperplanes}: We generated many random hyperplanes in the vector space. For each vector, we determined which side of each hyperplane it was on. We assigned a binary value (0 or 1) for its position relative to each plane. These binary values were then combined to form a binary string (\eg ``0110''), which served as a discrete scalar label for the vector.

\item \textbf{Sum of Distances to Reference Points}: We selected several random points in the vector space to act as reference points. For each vector, we calculated its distance to each of these reference points. We then summed these distances to produce a single continuous scalar value. This value was used as a continuous feature for the vector.

\end{itemize}

\stitle{Augmenting Scalar Datasets with Vector Data.}
We also created MHQ datasets from existing scalar datasets. 
% This approach ensures that the scalar data and its schema are complex and realistic. 
% We selected three tables from a IMDB dataset~\cite{maas2011learning}: Aka\_title, Aka\_name, and Title.
We selected 3 tables from IMDB dataset~\cite{maas2011learning} and 4 tables from TPCH dataset~\cite{tpch}.
These tables all contain columns with rich semantic information, such as movie titles or order comments. 
We processed these text-based columns using language models to generate vector embeddings. 
% For example, we converted each movie title in the title table into a corresponding vector. 
These embeddings were added to the original table as a vector column.
% , creating a large and realistic hybrid dataset. 
% In this way, we built benchmarks where the relationship between the scalar attributes and the generated vectors is naturally grounded in real-world data.
Through this approach, we developed benchmarks where scalar attributes and generated vectors maintain relationships that faithfully reflect real-world data.

\stitle{Query Generation.} For each table, we generated 1,000 hybrid queries involving a single vector column. For the Part and Aka\_title tables, which contain more than one semantically meaningful vector column, we additionally generated 1,000 MHQs that involve the weighted sum of distances across two vector columns. These settings ensure that both simple and multi-vector query patterns are adequately represented in our experimental evaluation.

Each hybrid query consists of two components: vector generation and scalar predicate generation. For the vector generation part, we first computed the value range of each vector dimension and then uniformly sampled random values within the corresponding range to form the query vector. For the scalar predicate generation, we randomly selected a subset of all scalar columns from the same table and assigned a predicate to each selected column. The predicate types include equality and range conditions, with operand values uniformly sampled from the domain of the corresponding column.

To ensure that the generated queries cover a wide range of realistic scenarios, we enforced the selectivity of scalar conditions to be uniformly distributed across 100 sub-intervals within the range [0,1]. 
% When the number of queries within a sub-interval exceeded 20, the queries in that interval were regenerated. 
When the number of queries within a sub-interval exceeded 20, queries that fall into this sub-interval need to be regenerated.
A similar procedure was applied to control the satisfaction rate of scalar predicates within the query vector’s neighborhood. For queries involving the weighted sum of distances over two vector columns, the weight $w_1$ was uniformly sampled from the interval [0,1], and the complementary weight was set as $w_2 = 1.0 - w_1$.

\section{Experiment}

Extensive experiments are conducted to validate \methodname{}. After outlining the experimental setup, the effectiveness and efficiency of the approach are assessed. Additionally, the impact of data updates is analyzed, alongside a black-box test of generalizability across various vector database systems. Finally, ablation studies are performed to isolate the contribution of individual components.

In our experiments, excluding the subsection 5.4, we adopted PostgreSQL (with the Pgvector~\cite{pgvector} extension) as the foundational database, paired with our proposed BoomHQ as the logical optimizer.
% This selection was motivated by the fact that this configuration offers the most comprehensive set of tunable parameters, whereas systems like Milvus~\cite{wang2021milvus} and OpenSearch~\cite{opensearch} do not support the configuration of $max\_scan\_tuples$ and $iterator\_scan$. 
This setup was chosen because it provides the widest range of tunable parameters, whereas Milvus~\cite{wang2021milvus} and OpenSearch~\cite{opensearch} do not support the configuration of $max\_scan\_tuples$ and $iterator\_scan$.
Furthermore, this setup directly supports both single-vector-column hybrid queries and multi-vector-column hybrid queries involving weighted distance computations.

\subsection{Experimental Setups}
\stitle{Datasets.} 
As shown in Table~\ref{tab:Datasets}, we use 11 datasets to evaluate \methodname{}. 
The Fungis dataset is a publicly available collection of fungal species images and metadata~\cite{HybridQueriesBenchmark}.
The Sift, Glove, and Deep1B datasets are sourced from the widely recognized ann-benchmarks~\cite{aumuller2020ann}. As these three datasets originally lack scalar attributes, we augmented them with scalar columns following the methodology detailed in Section 4 (donate as ``v $\rightarrow$ s'').
% Furthermore, we incorporate three tables—Aka\_title, Title, and Aka\_name—from the IMDb dataset~\cite{maas2011learning}. 
Furthermore, we incorporate three tables (Aka\_title, Title, and Aka\_name) from the IMDb dataset~\cite{maas2011learning}.
The remaining four tables (Part, Partsupp, Orders, and Lineitem) are derived from the TPC-H decision support benchmark~\cite{tpch}. The datasets originating from IMDb and TPC-H do not contain vector data. To address this, we generated vector embeddings from their semantically rich columns, using the process described in Section 4 (donate as ``s $\rightarrow$ v'').

\begin{table}[h]
\centering
  \caption{Overview of datasets}
  \label{tab:Datasets}
  % \begin{tabular}{l|c|c|c}
  \begin{tabular}{lccc}
    \toprule
    Benchmark & \#Type & \#Rows & \#Dimension\\
    \midrule
    \hline
    Fungis & v+s & 295,938 & 768 \\
    Sift & v$\rightarrow$s  & 1,000,000 & 128 \\
    Glove & v$\rightarrow$s  & 1,183,514 & 100 \\
    Deep1B & v$\rightarrow$s  & 9,990,000 & 96 \\
    Aka\_title & s$\rightarrow$v  & 361,472 & 768 \\
    Title & s$\rightarrow$v  & 2,528,312 & 768 \\
    Aka\_name & s$\rightarrow$v  & 901,343 & 768 \\
    Part & s$\rightarrow$v  & 200,000 & 768 \\
    Partsupp & s$\rightarrow$v  & 800,000 & 768 \\
    Orders & s$\rightarrow$v  & 1,500,000 & 768 \\
    Lineitem & s$\rightarrow$v  & 6,000,000 & 768 \\
  \bottomrule
\end{tabular}
\end{table}

\stitle{Baselines.} 
To conduct a evaluation of our proposed \methodname{} framework, we benchmark its performance in terms of both effectiveness and efficiency against 3 vector database systems: Pgvector~\cite{pgvector}, Milvus~\cite{wang2021milvus}, and OpenSearch~\cite{opensearch}.
Furthermore, to isolate and quantify the benefits of our logical optimization approach, we conduct a separate set of experiments where these 3 baseline systems are treated as black boxes. By applying our methodology to recommend optimal execution strategies and parameters  for these systems, we can directly measure the performance enhancements conferred by our logical optimizer, independent of the internal workings of the databases.

\stitle{Metrics.} 
To evaluate the effectiveness and efficiency of our proposed system for weighted multi-vector column hybrid queries, we employ the following two primary metrics.
The accuracy of our query system is quantified by the recall metric. 
% Recall measures the fraction of relevant items that are successfully retrieved by the system. In the context of our task, it is defined as the ratio of the number of correctly retrieved tuple IDs to the total number of actual relevant IDs in the ground truth set. A higher recall value indicates a more comprehensive and accurate search result.The formula for Recall is given by:
\begin{equation}
Recall = \frac{|S_{retrieved} \cap S_{groundtruth}|}{|S_{groundtruth}|}
\end{equation}
Where $S_{retrieved}$ represents the set of tuple IDs returned by our system in response to a query and $S_{groundtruth}$ represents the set of ground truth tuple IDs that are considered the correct results for that query.

The efficiency of our system are measured in terms of Queries Per Second (QPS). This metric evaluates the system's throughput by quantifying how many queries it can process within a one-second interval. 
% A higher QPS value signifies greater system efficiency and scalability.

The performance of vector search involves an inherent trade-off between recall and QPS, governed by mutable index parameters. 
To ensure a fair comparison of different execution strategies, we use the maximum achievable QPS at a fixed recall threshold (\eg recall@100 $\geq$ 0.90) as the primary performance metric.

\subsection{Evaluation of Effectiveness and Efficiency}

\begin{figure*}[t]
  \includegraphics[width=\textwidth]{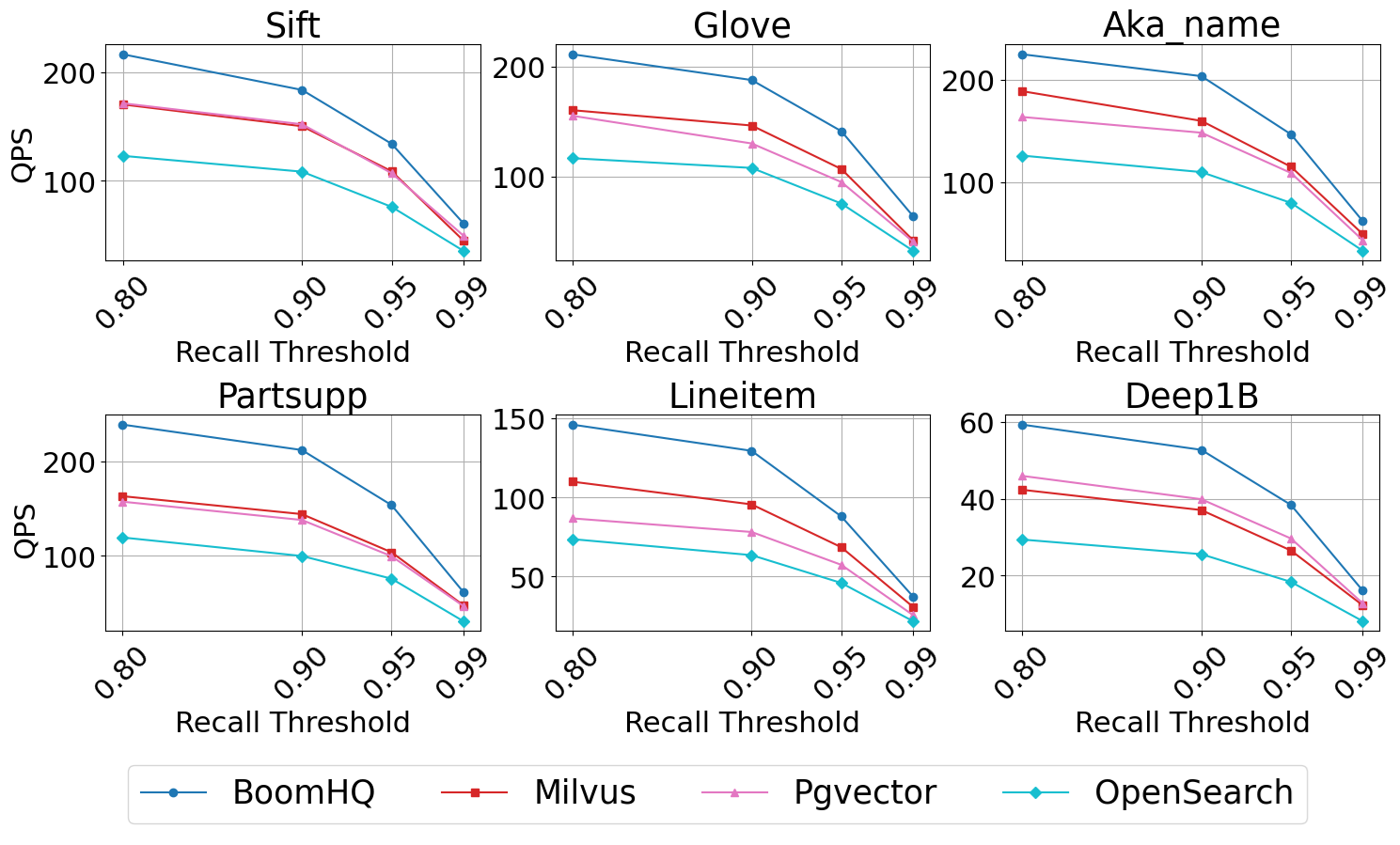}
  \caption{Single Vector Column Hybrid Query QPS vs. Recall Threshold Across Datasets }
  \label{fig:qps_comparison}
\end{figure*}

In this section, we present an evaluation of our proposed method. The evaluation is divided into two parts. First, we assess the performance of our execution strategy selection and parameter recommendation models for single-vector column hybrid queries. Second, we evaluate the extended framework, which incorporates a two-stage prediction model to support weighted multi-vector column hybrid queries.

We evaluate the effectiveness and efficiency of our execution parameter recommendation model by measuring the QPS in the varing recall threshold. 
% A meaningful analysis of QPS cannot be decoupled from the accuracy of the results. 
We evaluate the QPS under four fixed recall thresholds: 0.8, 0.9, 0.95, and 0.99. 
% This ensures a fair comparison of efficiency at specific levels of effectiveness. 
For brevity, we select 6 out of the 11 benchmarks for presentation.
As illustrated in  Figure~\ref{fig:qps_comparison}, our method consistently achieves a higher QPS than the baseline methods across all recall thresholds. 
Across 11 datasets, the QPS achieves an average improvement of 20\%, ranging from 8\% to 32\%.
This indicates that our model can find more efficient execution parameters to meet a given recall threshold and thus significantly improves query throughput.

\begin{figure}[t]
  \includegraphics[width=\columnwidth]{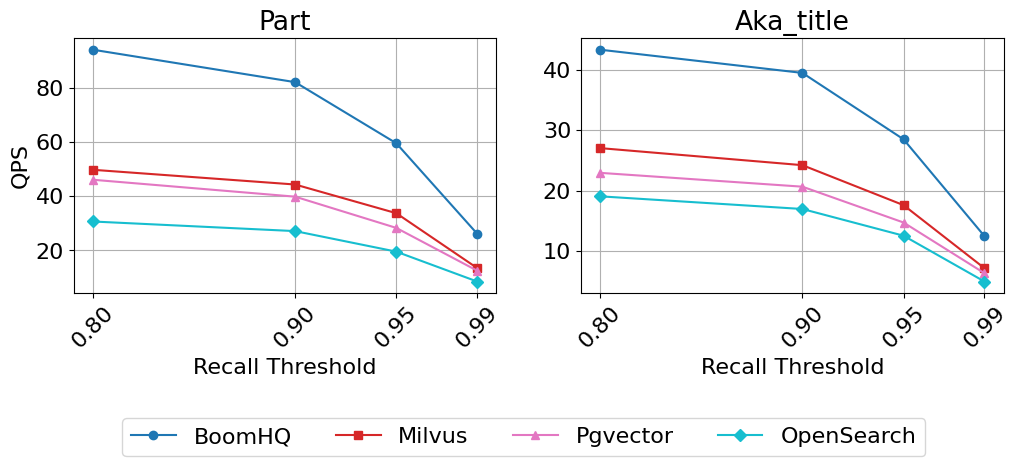}
  \caption{Weighted Multi-vector Column Hybrid Query QPS vs. Recall Threshold Across Datasets }
  \label{fig:multi_vector_effect}
\end{figure}

Building on the foundation of the single-vector optimization, we introduced a two-stage prediction model to support weighted multi-vector hybrid queries. 
% This enhancement is critical for handling more complex queries where vector columns have different degrees of importance.
We selected the two tables, namely Part and Aka\_title, and generated two vector columns on them.
Figure~\ref{fig:multi_vector_effect} illustrates that our method achieves average QPS improvements of 77\% (ranging from 66\% to 97\% for different recall thresholds) and 64\% (ranging from 59\% to 71\%) on the Part and Aka\_title datasets, respectively. 
Furthermore, on the Part dataset with a recall threshold of 0.80, \textbf{our average speedup reaches $2\times$, with more than 10 queries exhibiting a speedup exceeding $25\times$}.
% The chart shows a comparison of QPS between \methodname{} and other methods. 
Milvus and OpenSearch do not directly support queries on weighted multi-vector columns. 
% Therefore, in these two baselines, we rephrase the corresponding query statements using a two-stage approach.
Therefore, in these two baselines, we perform ANN search independently for each column and merge.
The results clearly demonstrate that our extended approach yields substantial gains in efficiency while maintaining high accuracy for complex hybrid queries.

\begin{figure}[t]
  \includegraphics[width=\columnwidth]{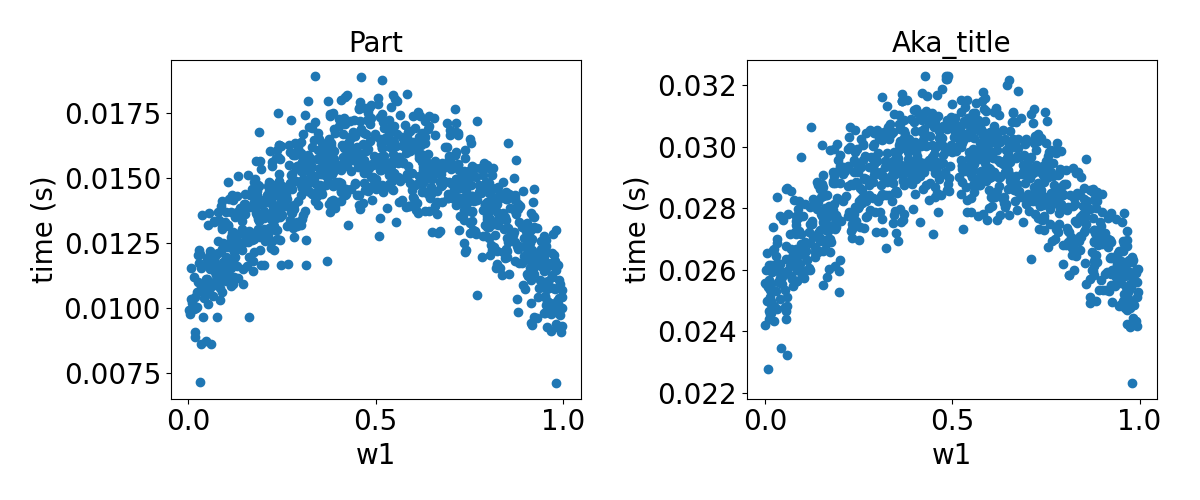}
  \caption{Effect of Parameter $w_1$ on Runtime with Recall Threshold 0.90 }
  \label{fig:time_w1}
\end{figure}

To provide deeper insight into our model's behavior in weighted multi-vector hybrid queries, we analyze the distribution of the weight $w_1$, which represents the importance of the first vector column in a hybrid query ($w_2=1.0-w_1$). 
Figure~\ref{fig:time_w1} illustrates the impact of $w_1$ on query latency across various queries, with the recall threshold set to 0.90.
From this figure, it can be observed that when the weight is heavily skewed, our method achieves superior optimization performance by automatically switching to a more appropriate index.
A similar phenomenon is also observed in our other recall threshold.
% This analysis reveals that our model effectively adjusts the query execution plan in response to the user-defined weights, prioritizing the more influential vector column to accelerate the search process. 
This analysis reveals that our model effectively adjusts the query execution plan in response to the user-defined weights and prioritizes the more influential vector column to accelerate the search process.
This adaptive behavior is key to the efficiency of our approach for weighted multi-vector hybrid queries.

\subsection{Impact of Data Updates}

This subsection extends the analysis to a dynamic environment. Here, we evaluate our method's performance during incremental data insertion.

Our experimental setup is designed to simulate continuous data insertion. 
% To create a challenging scenario that tests the model's capabilities, we ensure the newly inserted data has a skewed distribution compared to the original dataset. 
To create a challenging scenario, we make the newly inserted data follow a skewed distribution compared to the original dataset.
This directly evaluates the model's robustness against distribution shift. We measure the performance by varying the ratio of newly inserted data to the original dataset size. We established six distinct levels for this ratio: 0.005, 0.01, 0.05, 0.1, 0.5, and 1.0. A ratio of 1.0, for instance, means the amount of new data equals the entire original dataset size. 
% We assess the impact of these updates on two key metrics: query accuracy and query throughput.
Without loss of generality, we use the average QPS under a recall threshold of 0.9 as the evaluation metric.

\begin{figure}[t]
  \includegraphics[width=\columnwidth]{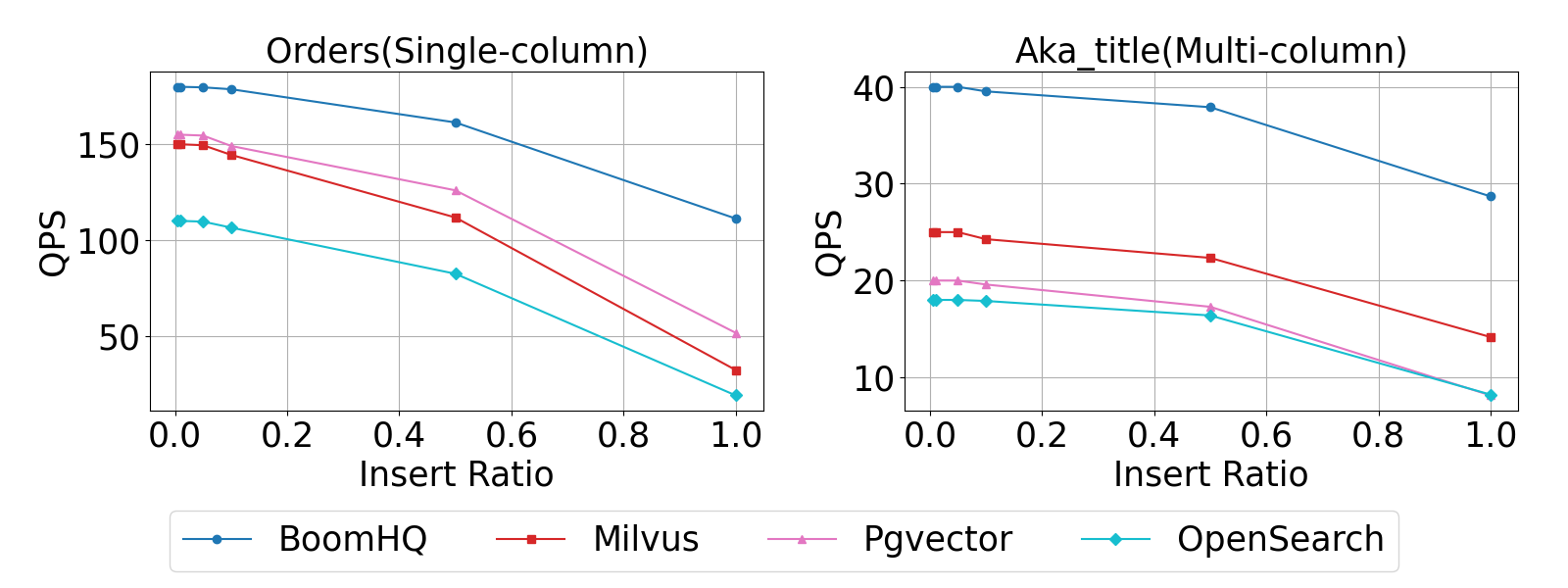}
  \caption{Effect of Insert Ratio on QPS with Recall Threshold 0.90 }
  \label{fig:qps_vs_insert_ratio}
\end{figure}

Figure~\ref{fig:qps_vs_insert_ratio} illustrates the average QPS as a function of the data insertion ratio. The results show that the QPS remains remarkably stable even with a significant influx of new data. There is only a minor degradation in performance when the volume of new data becomes comparable to the original dataset. 
We attribute this degradation to the increase in data volume.
As the decline rates of the curves corresponding to other methods are all higher than that of \methodname{}.

The results show that our autoencoder-based architecture successfully captures the evolving vector-scalar correlations introduced by newly inserted data. 
% \methodname{} demonstrates strong adaptability to database updates. 
This robustness ensures that \methodname{} is well-suited for dynamic, real-world application scenarios.

\subsection{Evaluation on Different Vector Databases}

Our methodology is designed to be database-agnostic. It functions by modeling the data and queries to select optimal execution strategies and recommend key parameters. This approach does not depend on the internal architecture of any specific database. To demonstrate this portability, we conducted experiments on two vector database systems: Milvus and OpenSearch. 
% These two database systems are treated as a black box, and their optimization is performed using external execution strategy and parameter control.
These two database systems are treated as a black box, and optimization is achieved by adjusting execution strategies and tuning system parameters.

Our evaluation focuses on weighted multi-vector column hybrid queries executed on the Part and Aka\_title tables. 
We compared the performance of a original system against our proposed optimization method. 
For the original system, the number of candidates for each vector column was set to $k' = \lambda k$, where $k$ is the number of results the original query expects to return. 
The parameters $\lambda$, $ef\_search$, and $iterator\_scan$ were applied consistently across all vector columns. 
These values were determined by selecting the best-performing combination from a grid search. 
In contrast, our optimized method employs a parameter recommendation model. 
This model individually recommends the optimal parameters for each separate column.

When applied to Milvus, the execution strategies and parameters recommended by our method resulted in an average QPS improvement ranging from 71\% to 93\% over the original version. 
Similarly, on OpenSearch, we observed a QPS improvement ranging from 85\% to 141\%. 
The results of our evaluation show that our method provides significant performance benefits on both systems.

\subsection{Ablation Study}

To validate the effectiveness and necessity of each component in our proposed method, we conducted an ablation study. We removed key components from our full model \methodname, and evaluated the performance of the resulting variants. We established a clear naming convention for these variants. For instance, the model without the data encoder is named \methodname~w.o.~DE, and the one without the query encoder is \methodname~w.o.~QE. Furthermore, we examined the sub-components in the query encoder. The variants without the query statistics encoding, the global selectivity estimation, and the local neighborhood probing are denoted as \methodname~w.o.~QE-Stats, \methodname~w.o.~QE-GSE, and \methodname~w.o.~QE-LNP, respectively.

\begin{figure*}[t]
  \includegraphics[width=\textwidth]{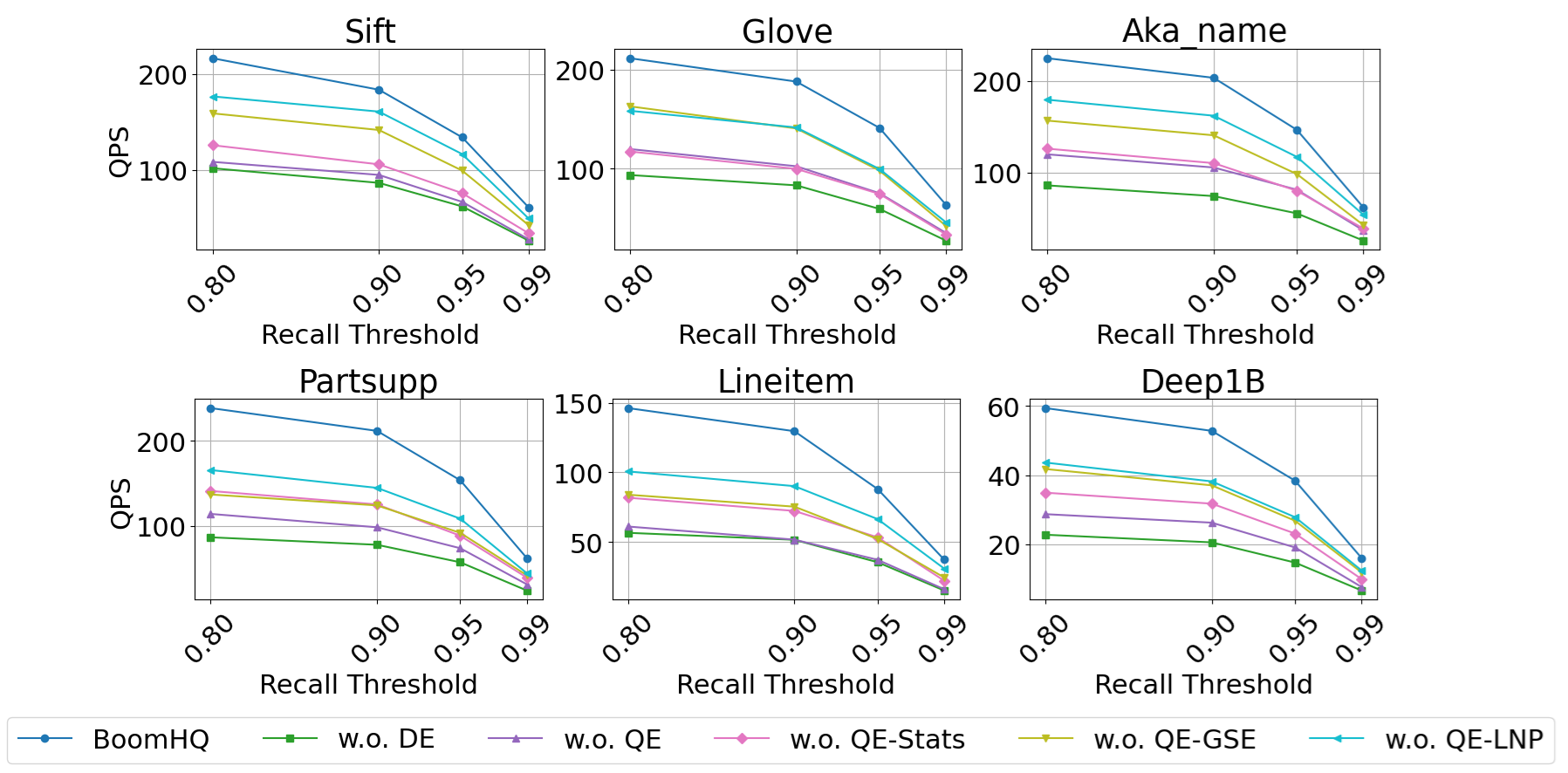}
  \caption{Ablation Study }
  \label{fig:ablation_study}
\end{figure*}

We first evaluated these model variants on the single-vector column hybrid query workload. 
As shown in Figure~\ref{fig:ablation_study}, when we removed the data encoder, the QPS of \methodname~w.o.~DE experienced a noticeable decline. This highlights the data encoder's crucial role in capturing the correlation between vectors and scalars effectively. 
Similarly, removing the entire query encoder in \methodname~w.o.~QE led to a sharp drop in QPS. 
This result confirms that a deep understanding of query feature is essential for achieving high performance. 
The degradation in performance for the sub-component variants further underscores the contribution of each specialized module in the query encoder.

Next, we conducted experiments using the more challenging weighted two-vector column hybrid queries to further assess the robustness of our design. 
The performance trends observed in this setting were consistent with our initial findings but even more pronounced. 
For example, the \methodname~w.o.~QE-GSE variant showed a significant decrease in QPS, which suggests that omitting the global selectivity estimation severely impairs the model's ability to generate efficient execute strategy under complex conditions. 
The performance gap between the full \methodname{} model and its ablated versions widened in this scenario. 
These results collectively demonstrate that all components are integral and indispensable. 
In conclusion, each module contributes synergistically to the final performance of \methodname.

\section{Conclusion}
In this paper, we proposed \methodname, a novel query rewriting framework. 
It is designed to recommend execution strategies and parameters for MHQs. 
Our method effectively uses the correlation between vector and scalar columns and also integrates global selectivity estimation and neighborhood pre-probing to inform the query rewrite process. 
The experimental results validate the effectiveness of \methodname. It significantly outperforms baseline methods in QPS under the given recall threshold. Furthermore, our method maintains stable and strong performance even as the data is continuously updated. 
As a logical optimizer, \methodname{} can be applied to optimize various vector databases, which highlights its practicality and broad applicability.

\section*{Funding}
This work was supported by NSFC (No. 62272008) and ZTE-PKU joint program.

\bibliography{sn-bibliography}% common bib file

\end{document}